\documentclass{article}

\usepackage{PRIMEarxiv}
\usepackage[utf8]{inputenc} 
\usepackage[T1]{fontenc}    
\usepackage{hyperref}       
\usepackage{url}            
\usepackage{booktabs}       
\usepackage{amsfonts}       
\usepackage{nicefrac}       
\usepackage{microtype}      
\usepackage{lipsum}
\usepackage{fancyhdr}       
\usepackage{graphicx}       
\graphicspath{{media/}}     
\usepackage{mathrsfs,mathtools,array}
\usepackage{booktabs,multicol}

\usepackage{subcaption}

\title{Monitoring multidimensional phenomena with a multicriteria composite performance interval approach}

\author{
  Ana Garcia-Bernabeu \\
  Department of Economy and Social Science \\
  Universitat Politècncia de València, Campus d'Alcoi \\
  Spain\\
  \texttt{angarber@upv.es} \\
   \And
  Adolfo Hilario-Caballero \\
  Institute of Control Systems and Industrial Computing (ai2)  \\
  Universitat Politècnica de Valencia \\
  Spain\\
  \texttt{ahilario@upv.es}
}

\begin{document}
\maketitle

\begin{abstract}
In the last two decades, composite indicators' construction to measure and compare multidimensional phenomena in a broad spectrum of domains has increased considerably. Different methodological approaches are used to summarize huge data sets of information in a single figure. This paper proposes a new approach that consists of computing a multicriteria composite performance interval based on different aggregation rules. The suggested approach provides an additional layer of information as the performance interval displays a lower bound from a non-compensability perspective, and an upper bound allowing for full-compensability. The outstanding features of this proposal are: (i) a distance-based multicriteria technique is taken as the baseline to construct the multicriteria performance interval (ii) the aggregation of distances/separation measures is made using particular cases of Minkowski's $L_p$ metrics; (iii) the span of the multicriteria performance interval can be considered as a sign of the dimensions or indicators balance.
\end{abstract}


\section{Introduction}
Composite indicators provide a one-dimensional metric to assess, monitor and predict the performance of complex phenomena approached from a multidimensional perspective such as human development, sustainability, innovation, or well being. According to \cite{saisana2002state}, a composite index involves the combination of a set of indicators that represents the different dimensions of the phenomenon to be measured.
Despite the fact that there is no single commonly accepted definition of a composite indicator, there is a common pattern in which the starting point is a complex phenomenon that includes different components that are to be compiled into a single indicator \cite{greco2017methodological}. Although the proliferation of composite indicators has grown exponentially over the last decade, there is no agreed consensus on an international standard for their construction.

The first attempt to establish a common guideline to construct a composite indicator targeted at  policy-makers, academics, the media and other interested parties was the ``Handbook of constructing composite indicators" \cite{nardo2008handbook}. In this manual,  a ten-step process from the development of a theoretical framework to the presentation and dissemination of a composite indicator is proposed to enhance the transparency and the soundness of the selected methodology. Main concerns around composite indicators are related with the lack of quality control and training for users, which can lead to send misleading and non-robust messages. Due to their relevance in policy-making decisions, the construction of composite indicators seems to be an important research issue from both theoretical and operational points of view \cite{munda2009noncompensatory}.

One of the most frequent criticism about construction of a composite indicator appears in  the  aggregation stage because this step defines the tool to add the criteria. The debate about the compensability, that appear when a deficit in one dimension can be offset by a surplus in another, supports the development of two groups: aggregators versus non-aggregators. Basically, an aggregation approach can be compensatory or non-compensatory depending on whether it permits compensability or not among indicators or dimensions \cite{tarabusi2013unbalance}. While the compensatory techniques deal with the imbalances of indicators and employ linear functions,  unbalance-adjusted functions are used in non-compensatory approaches.

Over the last few years, in multidimensional frameworks, when highly different dimensions should be aggregated, Multi-Criteria Decision Making (MCDM) methodologies have been claimed as highly suitable alternatives for constructing composite indicators \cite{samira2019}. Mainly, these methodologies have been widely used in selecting the  weights and in the aggregation stage. As to the multicriteria methods proposed to derive weights, we observe a high number of contributions applying data based methods such as Data Envelopment Analysis (DEA, \cite{charnes1978measuring}) as well as participatory based methods such as the Analytical Hierarchy Process (AHP, \cite{saaty1977scaling}). Regarding the aggregation stage, a significant group of MCDM methodologies adopt distance-based methods to construct composite indicators. 

This work aims to design a Multicriteria Composite Performance Interval (MCPI) instead of a single composite indicator looking at one of their technical weaknesses: the aggregation rule used for their construction. In our approach, we opted for the TOPSIS multicriteria tool, which is based on additive aggregation functions representing the distances to ideal and anti-ideal values. The proposal's novelty is to provide two indicators depending on the choice of the type of aggregation for the separation measures. The proposed methodology's added value implies extending the information provided by the classical relative clossenest to the ideal solution score.  Thus, the MCPI is composed of a lower and upper bound. The lower bound corresponds to non-compensability perspective, whereas the upper bound involves a full compensatory approach.   As an example of implementation and to demonstrate the advantages of using the proposed MCPI, it has been applied to measure the circular economy performance of EU countries on the basis of the structure of indicators provided by the EU monitoring framework. The proposed Circular Economy MCPI is developed to paint a comprehensive picture of the circular economy performance in 28 European countries and provide warning signals to policymakers on the areas where the dimensions need further improvements.

The paper is organized as follows. Section 2 reviews the compensability debate in constructing composite indicators and how some MCDM techniques have been applied to overcome this problem. In Section 3, the MCPI methodology is developed. In Section 4, an application of the proposed method to construct a Circular Economy MCPI is presented. Section 5 provides the conclusions as well as future lines of research.
\section{The compensability debate and MCDM approaches}

The selection of the aggregation procedure is one of the most discussed issues when constructing a composite indicator.  Based on the literature review of composite indicators, the most  widely applied aggregation procedures are linear aggregation rules implying complete substitutability among the various components considered. However, their applicability  depends on several strong theoretical and operational assumptions. The first one is the assumption of preferential independence, which in practice becomes very difficult to accomplish. The second one is related to the meaning of weights, which are viewed as substitution rates  instead of important coefficients. Despite these drawbacks, linear aggregation rules are a very intuitive and easy-to-use option.

A deep understanding of different aggregation rules in the framework of constructing composite indicators is provided in \cite{munda2009noncompensatory} and \cite{munda2012choosing}. The authors revised the debate on the use of aggregation rules by describing their relative pros and cons. Finally, they concluded that the use of non-linear/non-compensatory aggregation rules is advisable for reasons of theoretical consistency when weights are interpreted as importance coefficients or when the assumption of preferential independence does not hold. Under some conditions, these authors pointed out the benefits of using multicriteria aggregation procedures. While the compensatory techniques deal with the imbalances of indicators and employ linear functions,  unbalance-adjusted functions are used in non-compensatory approaches.

According to \cite{tarabusi2013unbalance}, in non-compensatory approaches the dimensions to be measured must be balanced and require the use of non-linear aggregation functions such as the geometric mean or the minimum  to penalize unbalance . The  Mazziota Pareto Index (MPI) introduced in 2007 \cite{mazziotta2007indicatore} is one of the first proposals to address the compensability  issue. Later, the same authors develop a newer variant of the previous method for spatio temporal comparisons known as Adjusted Mazziota Pareto Index (AMPI). In an attempt to overcome the problem of calculating a single figure, these authors have recently proposed a performance interval depending on the level of compensability of individual indicators \cite{Mazziotta2020}.  It should be noted that the previous work has inspired our proposal. However, while it relies on the power mean of order $r$ to deal with the compensability issue, we have introduced the use of performance intervals in the field of multicriteria decision making (MCDM) methodologies. 

Also in the literature of composite indicators, several authors claim that MCDM techniques are well suited for aggregating single indicators in a composite one in multidimensional frameworks (see \cite{saisana2002state, nardo2008handbook}). An overview of MCDM methodologies used to construct composite indicators was presented in \cite{samira2019}. On looking at the analysis made in the previous research, it can be remarkable the use of distance-based methodologies. Within this category, the reference point method \cite{wierzbicki1980use} was adapted by \cite{ruiz2011application}  to develop synthetic sustainability indicators dealing with different compensation degrees between the criteria and it has recently applied to construct composite indicators in a wide variety of domains (\cite{garcia2020multi},\cite{ruiz2021mrp}). 
These methods allow the decision-maker to define reference levels and reduce all indicators to a common scale using the so-called achievement scalarizing functions.

Other outranking multicriteria methodology where a full compensability prevails is the ``Technique for Ranking Preferences by Similarity to Ideal Solutions" TOPSIS initially introduced by \cite{yoon1981topsis}. This technique uses a compensatory aggregation rule of geometric distances to ideal and anti-ideal values. Some examples applying TOPSIS  to construct composite indicators can be found in \cite{wang2017smart},\cite{escrig2017measuring}, \cite{rosic2017method}, \cite{fu2020constructing}.

On the other hand, outranking related approaches such as ELECTRE (Elimination and Choice Expressing Reality, \cite{roy1968classement}) or PROMETHEE (Preference Ranking Organization Method for Enrichment Evaluation, \cite{brans1986select}) can avoid compensation thanks to the presence of veto threshold and ordinal comparison among alternatives. Some examples of application can be found in \cite{attardi2018non} to construct a Land-Use Policy Efficiency Index, or \cite{lopes2018regional} to assess regional tourism competitiveness.

In synthesis, despite the studies mentioned above, the debate on the compensability issue is continuing. Some recent studies emphasised the need to overcome linear aggregation rules in MCDM techniques by incorporating non-compensatory aggregation functions. However,  no previous work has dealt with integrating both perspectives by providing a multicriteria composite performance interval for different compensability degrees. Thus, by looking at the upper bound of the MCPI proposed in this paper,  we can get a big picture of the phenomenon under the compensability logic. In contrast, by looking at the lower bound, we can track the alternatives' specific weakness to be measured as it provides the worst performance by dimension or by indicator.

\section{Methodology:  Constructing the multicriteria composite performance interval (MCPI)}

In this paper, the TOPSIS multicriteria technique is proposed as a starting point to tackle the construction of a the MCPI.  The main feature of the proposed methodology is the choice of different compensability levels in the distance aggregation functions. In this way, the alternatives are classified according to an index that considers the ``shortest distance" from the ideal solution and the ``farthest distance" from the ``ideal-negative" solution. 

In general, TOPSIS \cite{tzeng1981multiple}  is implemented by the following stepwise procedure. After defining the decision matrix, including alternatives and criteria, the following step consists of normalizing the data. This is followed by computing the weighted normalized decision matrix.  Next, the positive and negative ideal solutions are identified to derive the separation measures. Thus,  the aggregation is made for each alternative's separation measures. Finally, the procedure ends by computing the relative closeness coefficient. The set of alternatives can be ranked according to the descending order of the closeness coefficient. In what follows, we extend the traditional formulation of this methodology, which in the earliest stages coincides with the steps for building a composite indicator to derive the proposed MCPI.

\subsection{Theoretical framework and data issues}

In our approach we consider a given theoretical framework from which the multidimensional phenomenon to be measured can be described in terms of alternatives, dimensions and indicators. From it,  the data are organized in a decision matrix $X =x_{ij}$  ($i=1, \ldots ,n$; $j=1,\ldots m$) where $n$ denotes the number of alternatives, and $m$ the number of indicators.

A previous data analysis should be taken to study the overall structure of the data set. At this stage the imputation of missing data and the application of multivariate analysis techniques are of great importance.

To allow comparability, the normalization of single indicators is a prior an necessary step as they are often expressed in different units of measurement. Starting from the initial data decision matrix, the normalized matrix $N = n_{ij}$ is constructed, where the normalized value of each indicator is obtained by applying the following transformation:

\begin{equation}
	n_{ij} =\frac{x_{ij}}{\sqrt{\sum\limits_{i=1}^n(x_{ij})^2}}
	\quad 
	(i=1, \ldots, n;\; j=1, \ldots ,m)
\end{equation}

\subsection{Weighting and distances to ideal and anti-ideal values}

Once the normalized matrix is obtained, the next step is the choice of weights and the aggregation method. The most common weighting and aggregation techniques rely on the simple arithmetic mean, which involves assigning equal weights to all indicators. There exist a rich menu of alternative weighting methods. Some of them are based on statistics such as Principal Component Analysis (PCA), Factor Analysis (FA). In other cases, subjective opinions from experts, citizens or politicians are taking into account to derive the weights through participatory methodologies such as the Benefit of the Doubt (BOD) or the Analytical Hierarchy Process (AHP) \cite{becker2017weights}.

As a starting point in our proposal, we have opted for the assignment of equal weights to build the weighted normalized matrix $V = v_{ij}$ as follows:

\begin{equation}
	v_{ij} = \omega_{j}\,n_{ij}    
\end{equation}


\noindent where $\omega^T = \omega_j = [\omega_1,\omega_2,\ldots,\omega_m]$, with $\sum\omega_j = 1$.

The positive and negative ideal solutions or the ideal and anti-ideal values, respectively are:

\begin{align*}
	V^+ = v_j^+ & = 
		\left\{
		\begin{array}{l}
			\displaystyle\max_{i}{v_{ij}\quad \forall j \in J_p}\\[2ex]
			\displaystyle\min_{i}{v_{ij}\quad \forall j \in J_n}
		\end{array}\right.\\[2ex]
	V^- = v_j^- & = 
		\left\{
		\begin{array}{l}
			\displaystyle\min_{i}{v_{ij}\quad \forall j \in J_p}\\[2ex]
			\displaystyle\max_{i}{v_{ij}\quad \forall j \in J_n}
		\end{array}\right.\\
\end{align*}

\noindent where $J_p$ is associated to benefit criteria  and  $J_n$ is associated to cost criteria. Therefore, the criteria separation measure to the ideal is computed for each criteria:

Notice that originally TOPSIS refer to the Euclidean distance to obtain the separation measures to the ideal and anti-ideal values. In fact, the Euclidean distance is a special case of Minkowski’s $L_p$ metric in an $n$-dimensional space (see \cite{berberian2012fundamentals}). 

\begin{equation}
L_p(x,y)=\left[\sum_{j=1}^{m}\lvert x_j - y_j\rvert^p\right]^{\textstyle\frac{1}{p}}
\end{equation}


\noindent where $p\geq1$. For $p=1$, we have the Manhattan distance involving and additive function and full compensability. When $p=2$  the Euclidean distance is obtained, and finally $p =\infty $ refers to the Tchebycheff distance which implies a non-compensatory approach.

\subsection{Defining the Multicriteria Composite Performance Interval}

At this stage, and to discuss how compensability among indicators can affect the overall ranking of alternatives, we propose to construct an interval instead of computing a single composite indicator based on the closeness to ideal solution index. The MCPI generates a lower and upper bound for the composite index.

\begin{itemize}

\item Lower bound:  The strong closeness to the ideal solution $C^S_i$. It involves a non-compensability choice in the aggregation stage of the separation measures. In this case an unbalance among indicators will have a negative effect on the value of the composite index.  When considering the separation measure to the ideal values, under strong perspective a penalty should be given to those criteria involving a maximum distance to the ideal and a minimum distance to the anti-ideal. In this case, we use the Tchebycheff's distances assuming a value of $p = +\infty$ to aggregate the positive distances and a value of $p = -\infty$ for negative distances respectively. 

\begin{equation}
\begin{aligned}
C^{S+}_{ik} & = L_{+\infty}(v_{ij},v^+_{j}) = \max_{j\in J_k}\, \lvert v_{ij} - v^+_{j}\rvert \\
C^{S-}_{ik} & = L_{-\infty}(v_{ij},v^+_{j}) = \min_{j\in J_k}\, \lvert v_{ij} - v^-_{j}\rvert
\end{aligned}
\end{equation}


\noindent where $J_k$ denotes the set of criteria belonging to dimension $k$, and $l_k$ is the number of criteria included.

Finally, we derive for each alternative the strong closeness to the ideal solution  $C^{S*}_{ik}$ as follows:

\begin{equation}
C^{S}_{ik}= \frac{C^{S-}_{ik}}{C^{S+}_{ik}+C^{S-}_{ik}}
\end{equation}

where $0\leq C^{S}_{ik} \leq 1$

\item Upper bound: The weak closeness to the ideal solution $C^{W}_{i}$. It corresponds to the upper bound of the multicriteria performance interval involving a full compensability in the aggregation stage of the separation measure. In this case an unbalance among indicators has any effect on the value of the composite index. In the $L_p$ metric, now we take the Manhattan distance with $p=1$, which corresponds to the arithmetic mean of the separation measures.

\begin{equation}
\begin{aligned}
C^{W+}_{ik} & =  L_{1}(v_{ij},v^+_{j}) = \frac{1}{l_k} \sum_{j\in J_k} \lvert v_{ij} - v^+_{j}\rvert \\
C^{W-}_{ik} & = L_{1}(v_{ij},v^-_{j}) = \frac{1}{l_k} \sum_{j\in J_k}\lvert v_{ij} - v^-_{j}\rvert
\end{aligned}
\end{equation}


Analogously, we derive the weak closeness to the ideal solution  $C^{W*}_{ik}$ as follows:

\begin{equation}
C^{W}_{ik}= \frac{C^{W-}_{ik}}{C^{W+}_{ik}+C^{W-}_{ik}}
\end{equation}

\end{itemize}
 
\bigskip
Thus, the MCPI for each dimension, $\delta_{ik}$, takes the following form:

\begin{equation}\label{eq:MCPI_dim}
	\delta_{ik} = \left[C^{S}_{ik}, C^{W}_{ik}\right]
\end{equation}

Once, the aggregation rule is made for the first level, namely, from indicators to dimensions, the modeller has several options to derive the overall MCPI when the dimensions should be aggregated. In our case, we propose a lower bound which takes the minimum value of the dimensions, and a upper bound corresponding to the mean of the weak closeness to the ideal solution. 

\begin{equation}
	\gamma_{i} = \left[\min_{k}{C^{W}_{ik}},\frac{1}{l} \sum_{k=1}^{l} {C^{W}_{ik}}\right]
\end{equation}

The question now is, how the user of the MCPI should rank the alternatives? The answer is not trivial and depends on the subject under study and the modeller's objectives. A quite sound option is to rank the alternative by the value of the upper limit, but looking at the span of the MCPI, which  can be considered as a sign of the unbalanced. For example, looking at the $\delta_{ik}$, the greater the span of the interval, the greater the imbalance of the single indicators, whereas the unbalance between dimensions comes from the length of the $\gamma_{i}$.

Therefore, policy-makers setting targets to improve a given multidimensional phenomenon could prioritise which alternative and in which dimensions to act on by looking at the span of the MCPI. Based on the balance property of the MCPI, the alternatives are assigned a Balance Rating ($\beta$--Rating) according to the following scale. 

\begin{table}[h]
	\centering
	\caption{MCPI interval balance rating}
	\begin{tabular}{lc}
	\toprule
	Category span	& $\beta$--Rating icon\\
	\midrule
	High balance MCPI & $\star\star\star$\\
	Medium balance MCPI & $\star\star$\\
	Low balance MCPI & $\star$\\		
	\bottomrule
	\end{tabular}
\end{table}

\begin{itemize}
\item Alternatives with a MCPI span ranging from $0\,\%$ to $33.33\,\%$ of maximum span can receive a high \mbox{$\beta$--Rating} icon (3 stars).
\item Alternatives with a MCPI span ranging from $33.33\,\%$ to $66.67\,\%$ of maximum span can receive a medium \mbox{$\beta$--Rating} icon (2 stars).
\item Alternatives with a MCPI span ranging above the $66.67\,\%$ of maximum span can receive a low \mbox{$\beta$--Rating} icon (1 star).
\end{itemize}

\section{Case study: the construction of the Circular Economy- MCPI}

This section applies the MCPI approach following a set of Circular Economic indicators with data extracted from the EU Circular Economy monitoring framework \cite{Europa2019}. Since 2018, and in line with the highlighted relevance of sustainability concerns in Europe, a monitoring framework to assess Circular Economic issues' performance was presented by the European Commission. This is an instrument for monitoring key trends in the transition towards a more circular economy model in Europe, which makes it possible to assess whether the measures put in place and the involvement of all stakeholders have been sufficiently effective and identify best practices in the Member States. An initial proposal to provide a composite indicator for EU member states by aggregating a different set of indicators was presented in \cite{garcia2020process}. 

\subsection{Circular Economy monitoring framework, indicators and data}

 The circularity assessment of EU countries is accomplished according to the Circular Economy monitoring framework, which departs from ten key indicators including other sub-indicators and grouped in the following four broad dimensions: production and consumption, waste management, secondary raw materials, and competitiveness and innovation. The indicators' data come from Eurostat, the Joint Research Centre, and the European Patent Office. To evaluate EU countries' progress towards circular economy, the information is disseminated throughout tables and graphs for cross-national comparison in the following web page \url{https://ec.europsa.eu/eurostat/web/circular-economy/indicators/ monitoring-framework}.

From this monitoring framework, we select ten sub-indicators grouped into the four dimensions and aspects of the circular economy. A descriptive analysis of the selected indicators is provided in Table \ref{tab:tab_01}. In our application, a total of 280 observations are available for the 28 EU countries and ten indicators. The indicators with the highest averages relate to the amount of waste per capita measured in kilograms, waste per unit of GDP and trade in secondary raw materials measured in tonnes. It can be seen that the data, when considering all 28 countries, show little homogeneity, as indicated by the kurtosis coefficient, especially in the mean of the amount of waste per unit of GDP, the number of patents and the contribution of recycled raw materials to the demand for raw materials. This result is indicative of the heterogeneity of the data set.

\begin{table*}[htbp]
  \centering
  \caption{Selected indicators of Circular Economy and data issues. Source European Commission Monitoring Framework  (2018)}
  \resizebox{\textwidth}{!}{
  \begin{tabular}{rp{18.5em}p{8.335em}rrrrr}
      \toprule
            & Dimension/ Indicator & Data Source & {Mean} & {st} & {Max} & {Min} & {Kurtosis} \\
      \midrule
            & {Production and consumption (PC)} &  &       &       &       &       &  \\
      \midrule     
      1a    & Generation of municipal waste per capita (Kg per capita) & European Statistical System & 497.64 & 128.24 & 814.00 & 272.00 & 0.89 \\
      1b    & Generation of waste excluding major mineral wastes per GDP unit (Kg per thousand euro, chain linked volumes (2010)) & European Statistical System & 105.21 & 133.90 & 646.00 & 27.00 & 11.61 \\
      1c    & Generation of waste excluding major mineral wastes per domestic material consumption (percentage) & European Statistical System & 12.51 & 7.47  & 30.50 & 4.80  & 0.77 \\
     \midrule 
	& {Waste Management (WM)} &  &       &       &       &       &  \\
	\midrule     
      2a    & Recycling rate of municipal waste (percentage) & European Statistical System & 38.69 & 14.81 & 67.10 & 10.00 & -0.57 \\
	  2b    & Recycling rate of all waste excluding major mineral waste (percentage) & European Statistical System & 49.43 & 18.12 & 80.00 & 10.00 & -0.06 \\
	  \midrule     
	& {Secondary Raw Materials (SRM)} & &       &       &       &       &  \\
	  \midrule     
      3a    &{Contribution of recycled material to raw materials demand} & European Statistical System & 9.26  & 6.75  & 29.00 & 1.50  & 1.43 \\
3b    & Trade in recyclable raw materials (tonnes) & European Statistical System & 3437406.64 & 4323147.25 & 14100540.00 & 59472.00 & 0.92 \\
\midrule     
& {Competitiveness and Innovation (CI)} & \ &       &       &       &       &  \\
\midrule     
      4a    & Persons employed (percentage of total employment) & European Statistical System & 1.82  & 0.46  & 2.82  & 1.10  & 0.28 \\
4b    & Value added at factor cost (percentage of GDP at current prices) & European Statistical System & 0.97  & 0.24  & 1.30  & 0.36  & 1.41 \\
4c    & Number of patents related to recycling and secondary raw materials & European Patent Office & 0.59  & 0.64  & 2.58  & 0.00  & 2.40 \\
      \bottomrule
      \end{tabular}}%
  \label{tab:tab_01}%
\end{table*}%

\subsection{The Circular Economy -- MCPI by dimensions}

Table \ref{tab:dimensions} shows the results of the MCPI  by country and for the four dimensions of production and consumption $\delta_{PC}$, waste management $\delta_{WN}$, secondary raw materials $\delta_{SRM}$, and competitiveness and innovation $\delta_{CI}$. When the information is presented by dimension, the particular performance of each country in each of the areas could be analysed. By applying a compensatory approach to the indicators in each area, the value of the upper limit of the interval is obtained. Under a non-compensatory perspective, the lower limit indicates the value of the worst performing indicator in each dimension. On the other hand, the length of the MCPI provides information on the balance or imbalance of the indicators. For example, if we look at the case of Austria, we can see how the information displayed in the waste management dimension $\left[80.0, 81.1\right]$ presents a high value at both ends of the interval indicating that all indicators of this dimension are balanced. However, we can observe that for the dimensions of production and consumption $\left[43.4, 83.2\right]$ and innovation and competitiveness $\left[8.1, 30.7\right]$ the length of the MPCI is considerable and therefore indicative that some indicators in these dimensions perform too poorly. From a policy perspective, attention should be paid to those dimensions either where the value is too low or where there is an imbalance.

\begin{table*}[t]
	\centering
	\caption{The MCPI of Circular Economy by dimension in Europe (year 2018)}
	{
	\begin{tabular}{ l r@{\hspace{2pt}}  r@{\hspace{4pt}}  r  r@{\hspace{2pt}}  r@{\hspace{4pt}}  r  r@{\hspace{2pt}}  r@{\hspace{4pt}}  r r@{\hspace{2pt}}  r@{\hspace{4pt}}  r }

\toprule

 &  \multicolumn{3}{c}{$\delta_\mathrm{PC}\;(\%)$} & \multicolumn{3}{c}{$\delta_\mathrm{WM}\;(\%)$} & \multicolumn{3}{c}{$\delta_\mathrm{SRM}\;(\%)$}  & \multicolumn{3}{c}{$\delta_\mathrm{CI}\;(\%)$} \\

\midrule

Austria	&	[	&	43.4,	&	83.2\hspace{2pt}]	&	[	&	80.0,	&	81.8\hspace{2pt}]	&	[	&	23.4,	&	29.5\hspace{2pt}]	&	[	&	8.1,	&	30.7\hspace{2pt}] \\
Belgium	&	[	&	4.7,	&	63.4\hspace{2pt}]	&	[	&	77.8,	&	87.3\hspace{2pt}]	&	[	&	72.6,	&	73.8\hspace{2pt}]	&	[	&	0.0,	&	37.5\hspace{2pt}] \\
Bulgaria	&	[	&	23.0,	&	44.2\hspace{2pt}]	&	[	&	24.3,	&	31.1\hspace{2pt}]	&	[	&	3.5,	&	4.1\hspace{2pt}]	&	[	&	0.0,	&	24.6\hspace{2pt}] \\
Croatia	&	[	&	70.5,	&	86.3\hspace{2pt}]	&	[	&	26.8,	&	43.1\hspace{2pt}]	&	[	&	3.0,	&	7.5\hspace{2pt}]	&	[	&	15.5,	&	40.5\hspace{2pt}] \\
Cyprus	&	[	&	32.7,	&	87.7\hspace{2pt}]	&	[	&	10.7,	&	20.2\hspace{2pt}]	&	[	&	0.0,	&	2.1\hspace{2pt}]	&	[	&	0.0,	&	22.8\hspace{2pt}] \\
Czechia	&	[	&	77.5,	&	90.2\hspace{2pt}]	&	[	&	42.9,	&	56.9\hspace{2pt}]	&	[	&	6.2,	&	14.7\hspace{2pt}]	&	[	&	9.3,	&	37.1\hspace{2pt}] \\
Denmark	&	[	&	0.0,	&	80.0\hspace{2pt}]	&	[	&	69.9,	&	71.3\hspace{2pt}]	&	[	&	9.6,	&	16.7\hspace{2pt}]	&	[	&	7.1,	&	35.6\hspace{2pt}] \\
Estonia	&	[	&	0.0,	&	12.2\hspace{2pt}]	&	[	&	0.0,	&	16.1\hspace{2pt}]	&	[	&	4.0,	&	23.7\hspace{2pt}]	&	[	&	28.3,	&	61.8\hspace{2pt}] \\
Finland	&	[	&	48.5,	&	85.0\hspace{2pt}]	&	[	&	38.6,	&	47.7\hspace{2pt}]	&	[	&	2.3,	&	8.9\hspace{2pt}]	&	[	&	24.7,	&	61.7\hspace{2pt}] \\
France	&	[	&	48.6,	&	81.6\hspace{2pt}]	&	[	&	61.1,	&	62.1\hspace{2pt}]	&	[	&	36.6,	&	50.8\hspace{2pt}]	&	[	&	10.8,	&	31.4\hspace{2pt}] \\
Germany	&	[	&	38.4,	&	79.5\hspace{2pt}]	&	[	&	61.4,	&	81.1\hspace{2pt}]	&	[	&	38.2,	&	70.1\hspace{2pt}]	&	[	&	9.2,	&	36.7\hspace{2pt}] \\
Greece	&	[	&	49.5,	&	78.3\hspace{2pt}]	&	[	&	17.7,	&	23.7\hspace{2pt}]	&	[	&	6.4,	&	8.4\hspace{2pt}]	&	[	&	0.0,	&	6.8\hspace{2pt}] \\
Hungary	&	[	&	70.3,	&	88.9\hspace{2pt}]	&	[	&	46.6,	&	47.6\hspace{2pt}]	&	[	&	5.1,	&	12.3\hspace{2pt}]	&	[	&	13.9,	&	30.0\hspace{2pt}] \\
Ireland	&	[	&	39.9,	&	88.1\hspace{2pt}]	&	[	&	44.3,	&	46.4\hspace{2pt}]	&	[	&	0.3,	&	1.7\hspace{2pt}]	&	[	&	9.3,	&	30.9\hspace{2pt}] \\
Italy	&	[	&	30.0,	&	70.3\hspace{2pt}]	&	[	&	69.7,	&	76.2\hspace{2pt}]	&	[	&	61.9,	&	63.2\hspace{2pt}]	&	[	&	8.9,	&	30.7\hspace{2pt}] \\
Latvia	&	[	&	75.1,	&	93.1\hspace{2pt}]	&	[	&	0.0,	&	13.6\hspace{2pt}]	&	[	&	4.3,	&	8.1\hspace{2pt}]	&	[	&	19.8,	&	46.1\hspace{2pt}] \\
Lithuania	&	[	&	59.5,	&	83.8\hspace{2pt}]	&	[	&	74.4,	&	78.6\hspace{2pt}]	&	[	&	7.8,	&	8.9\hspace{2pt}]	&	[	&	0.0,	&	34.0\hspace{2pt}] \\
Luxembourg	&	[	&	2.0,	&	78.6\hspace{2pt}]	&	[	&	68.3,	&	72.6\hspace{2pt}]	&	[	&	17.7,	&	25.5\hspace{2pt}]	&	[	&	100.0,	&	100.0\hspace{2pt}] \\
Malta	&	[	&	27.9,	&	82.6\hspace{2pt}]	&	[	&	0.0,	&	23.1\hspace{2pt}]	&	[	&	0.8,	&	12.0\hspace{2pt}]	&	[	&	0.0,	&	0.0\hspace{2pt}] \\
Netherlands	&	[	&	10.1,	&	65.1\hspace{2pt}]	&	[	&	80.4,	&	84.4\hspace{2pt}]	&	[	&	88.1,	&	93.9\hspace{2pt}]	&	[	&	2.4,	&	32.9\hspace{2pt}] \\
Poland	&	[	&	53.7,	&	79.4\hspace{2pt}]	&	[	&	41.7,	&	53.5\hspace{2pt}]	&	[	&	17.2,	&	23.3\hspace{2pt}]	&	[	&	26.6,	&	56.0\hspace{2pt}] \\
Portugal	&	[	&	56.6,	&	85.8\hspace{2pt}]	&	[	&	33.5,	&	46.5\hspace{2pt}]	&	[	&	2.2,	&	8.5\hspace{2pt}]	&	[	&	0.0,	&	17.0\hspace{2pt}] \\
Romania	&	[	&	64.0,	&	90.9\hspace{2pt}]	&	[	&	1.9,	&	15.0\hspace{2pt}]	&	[	&	0.0,	&	4.0\hspace{2pt}]	&	[	&	5.8,	&	17.3\hspace{2pt}] \\
Slovakia	&	[	&	62.1,	&	81.6\hspace{2pt}]	&	[	&	46.1,	&	47.3\hspace{2pt}]	&	[	&	3.5,	&	8.0\hspace{2pt}]	&	[	&	0.8,	&	16.8\hspace{2pt}] \\
Slovenia	&	[	&	60.5,	&	83.6\hspace{2pt}]	&	[	&	85.6,	&	92.7\hspace{2pt}]	&	[	&	7.9,	&	19.1\hspace{2pt}]	&	[	&	0.0,	&	29.9\hspace{2pt}] \\
Spain	&	[	&	42.6,	&	76.9\hspace{2pt}]	&	[	&	43.4,	&	47.4\hspace{2pt}]	&	[	&	29.4,	&	40.1\hspace{2pt}]	&	[	&	18.1,	&	39.8\hspace{2pt}] \\
Sweden	&	[	&	70.1,	&	89.1\hspace{2pt}]	&	[	&	55.7,	&	59.3\hspace{2pt}]	&	[	&	16.2,	&	17.5\hspace{2pt}]	&	[	&	9.6,	&	28.0\hspace{2pt}] \\
United Kingdom	&	[	&	31.1,	&	72.7\hspace{2pt}]	&	[	&	59.7,	&	64.1\hspace{2pt}]	&	[	&	53.8,	&	73.5\hspace{2pt}]	&	[	&	8.6,	&	29.9\hspace{2pt}] \\

\bottomrule

\end{tabular}

	}
	\label{tab:dimensions}
\end{table*}

\subsection{The overall Circular Economy -- MCPI}

To assess the overall performance in terms of EU Member states' circular economy, we construct the Circular Economy -- MCPI. Table \ref{tab:overall} sorts the EU countries by the upper bound of the $\gamma_{i}$--MCPI . Moreover, we add a column to highlight the span of the MCPI. In the last column, we highlight those countries with more balance scores by the $\beta$--rating scale. Luxembourg, Netherlands and Germany head up the ranking with the highest scores of the upper bound of the Circular Economy -- MCPI. In contrast, Greece, Estonia and Bulgaria occupy the lowest positions in the ranking. Notice that, for Luxembourg, the best-positioned country, the span of the MCPI is the longest, which is indicative of a significant imbalance in one of the dimensions and for this reason it receive a $\beta$--Rating of 1 star ($\star$).  Then, we need to look at the information provided by dimensions  in Table \ref{tab:dimensions}. It can be seen that, a great imbalance in the production and consumption dimension as highlighted in  $\delta_{PC} = \left[2.0, 78.6\right]$, which is compensated by the high performance of the competitiveness and innovation dimension $\delta_{CI}=\left[100.0, 100.0\right]$.

\begin{table*}[h]
	\centering
	\caption{Multicriteria composite performance interval of Circular Economy in Europe (year 2018)}
	{
	\begin{tabular}{ l c r@{\hspace{2pt}}  r@{\hspace{4pt}} r r c}

\toprule

 & Rank & \multicolumn{3}{c}{$\gamma_i$} & Span & $\beta$--rating \\

\midrule

Luxembourg	&	1	&	[	&	25.5,	&	69.2\hspace{2pt}]	&	43.7	&	$\star$ \\
Netherlands	&	2	&	[	&	32.9,	&	69.1\hspace{2pt}]	&	36.2	&	$\star$ \\
Germany	&	3	&	[	&	36.7,	&	66.8\hspace{2pt}]	&	30.2	&	$\star\star$ \\
Belgium	&	4	&	[	&	37.5,	&	65.5\hspace{2pt}]	&	28.0	&	$\star\star$ \\
Italy	&	5	&	[	&	30.7,	&	60.1\hspace{2pt}]	&	29.4	&	$\star\star$ \\
United Kingdom	&	6	&	[	&	29.9,	&	60.0\hspace{2pt}]	&	30.2	&	$\star\star$ \\
France	&	7	&	[	&	31.4,	&	56.5\hspace{2pt}]	&	25.1	&	$\star\star$ \\
Slovenia	&	8	&	[	&	19.1,	&	56.3\hspace{2pt}]	&	37.3	&	$\star$ \\
Austria	&	9	&	[	&	29.5,	&	56.3\hspace{2pt}]	&	26.8	&	$\star\star$ \\
Poland	&	10	&	[	&	23.3,	&	53.1\hspace{2pt}]	&	29.7	&	$\star\star$ \\
Lithuania	&	11	&	[	&	8.9,	&	51.3\hspace{2pt}]	&	42.4	&	$\star$ \\
Spain	&	12	&	[	&	39.8,	&	51.0\hspace{2pt}]	&	11.2	&	$\star\star\star$ \\
Denmark	&	13	&	[	&	16.7,	&	50.9\hspace{2pt}]	&	34.2	&	$\star$ \\
Finland	&	14	&	[	&	8.9,	&	50.8\hspace{2pt}]	&	41.9	&	$\star$ \\
Czechia	&	15	&	[	&	14.7,	&	49.7\hspace{2pt}]	&	35.1	&	$\star$ \\
Sweden	&	16	&	[	&	17.5,	&	48.5\hspace{2pt}]	&	30.9	&	$\star\star$ \\
Hungary	&	17	&	[	&	12.3,	&	44.7\hspace{2pt}]	&	32.4	&	$\star\star$ \\
Croatia	&	18	&	[	&	7.5,	&	44.4\hspace{2pt}]	&	36.8	&	$\star$ \\
Ireland	&	19	&	[	&	1.7,	&	41.8\hspace{2pt}]	&	40.1	&	$\star$ \\
Latvia	&	20	&	[	&	8.1,	&	40.2\hspace{2pt}]	&	32.2	&	$\star\star$ \\
Portugal	&	21	&	[	&	8.5,	&	39.4\hspace{2pt}]	&	31.0	&	$\star\star$ \\
Slovakia	&	22	&	[	&	8.0,	&	38.4\hspace{2pt}]	&	30.4	&	$\star\star$ \\
Cyprus	&	23	&	[	&	2.1,	&	33.2\hspace{2pt}]	&	31.1	&	$\star\star$ \\
Romania	&	24	&	[	&	4.0,	&	31.8\hspace{2pt}]	&	27.8	&	$\star\star$ \\
Malta	&	25	&	[	&	0.0,	&	29.4\hspace{2pt}]	&	29.4	&	$\star\star$ \\
Greece	&	26	&	[	&	6.8,	&	29.3\hspace{2pt}]	&	22.5	&	$\star\star$ \\
Estonia	&	27	&	[	&	12.2,	&	28.5\hspace{2pt}]	&	16.2	&	$\star\star\star$ \\
Bulgaria	&	28	&	[	&	4.1,	&	26.0\hspace{2pt}]	&	21.9	&	$\star\star\star$ \\

\bottomrule

\end{tabular}

	}
	\label{tab:overall}
\end{table*}

Once we have obtained an overall assessment of countries' circular economy performance, we have noticed that the fourth dimension (CI) has a significant contribution to the final ranking. This means that a country like Luxembourg can head the ranking with the best performance in this dimension. For this reason, we have profiled circular economy performance by eliminating this dimension, namely, taking into account the dimensions that originally characterise the circular economy (production and consumption, waste management and secondary raw materials). Table \ref{tab:overall_but_ci}  display the MCPI but excluding the competitiveness and innovation dimension. It also provides the rank position, and its variation for the overall ranking. We observe how Luxembourg now is downgraded eight places and, finally, how Netherlands, Germany and Belgium top the ranking. Moreover, we can also see that they are more balanced as shown by the $\beta$--rating. 

\begin{figure*}[h]
	\centering
	\begin{subfigure}{0.475\textwidth}
		\resizebox{\columnwidth}{!}{\includegraphics{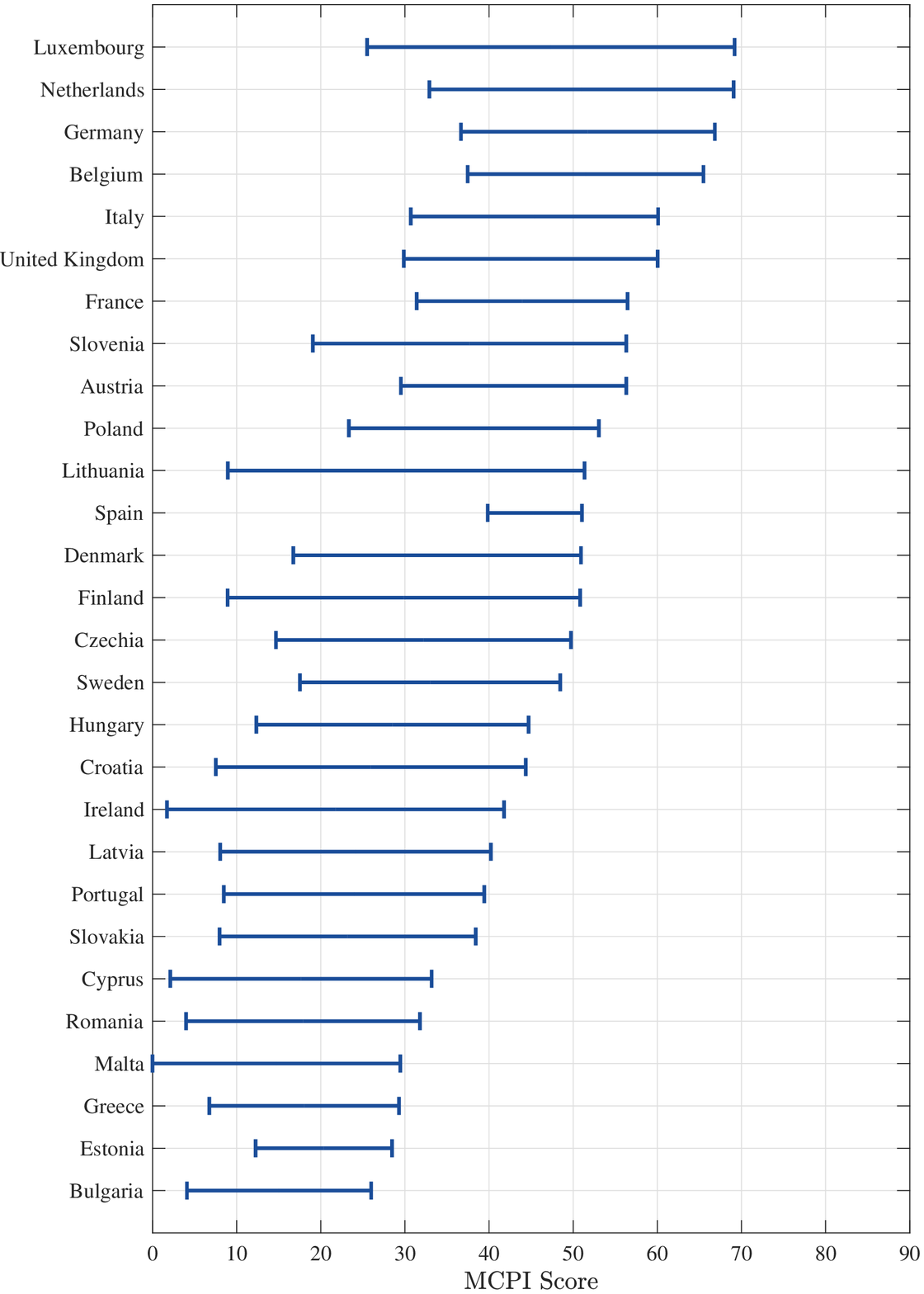}}
		\caption{All four dimensions}
		\label{fig:MCPI}			
	\end{subfigure}
	\hfill
	\begin{subfigure}{0.475\textwidth}
		\resizebox{\columnwidth}{!}{\includegraphics{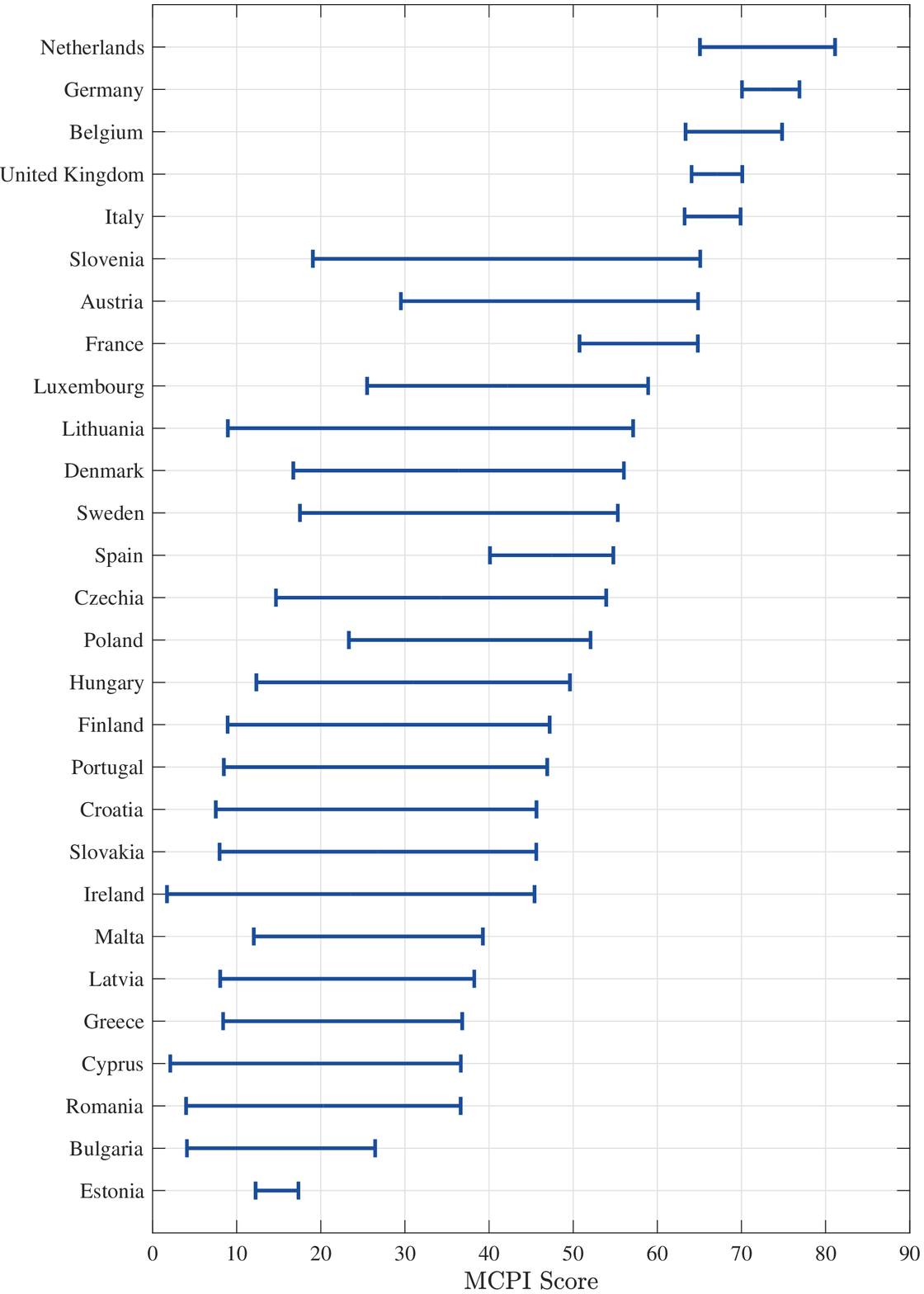}}
		\caption{Without CI dimension}
		\label{fig:MCPI_but_CI}			
	\end{subfigure}
	\caption{Overall Circular Economy Multicriteria Composite Performance Interval (MCPI)}
	\label{fig:MCPI:completa}	
\end{figure*}

Figure \ref{fig:MCPI:completa} graphically shows the compared MCPI scores when including all four dimensions and without the competitiveness and innovation dimension. As can be seen, in the later case, the heading countries presents a shorter MCPI than when including all dimensions.

It should be noted that the Circular Economy MCPI within dimensions and for the overall performance provides a much richer information than a single composite index allowing the modeller to interpret the meaning of the composite measures and carry out corrective measures where necessary. 

\begin{table*}[h]
	\centering
	\caption{Multicriteria composite performance interval of Circular Economy in Europe excluding CI dimension (year 2018)}
	{
	\begin{tabular}{ l c >{$}c<{$} r@{\hspace{2pt}}  r@{\hspace{4pt}} r r c}

\toprule

 & Rank & \Delta\mathrm{Rank} & \multicolumn{3}{c}{$\gamma_i$} & Span & $\beta$--rating \\

\midrule

Netherlands	&	1	&	+1	&	[	&	65.1,	&	81.1\hspace{2pt}]	&	16.1	&	$\star\star\star$ \\
Germany	&	2	&	+1	&	[	&	70.1,	&	76.9\hspace{2pt}]	&	6.8	&	$\star\star\star$ \\
Belgium	&	3	&	+1	&	[	&	63.4,	&	74.8\hspace{2pt}]	&	11.5	&	$\star\star\star$ \\
United Kingdom	&	4	&	+2	&	[	&	64.1,	&	70.1\hspace{2pt}]	&	6.0	&	$\star\star\star$ \\
Italy	&	5	&	+0	&	[	&	63.2,	&	69.9\hspace{2pt}]	&	6.7	&	$\star\star\star$ \\
Slovenia	&	6	&	+2	&	[	&	19.1,	&	65.1\hspace{2pt}]	&	46.1	&	$\star$ \\
Austria	&	7	&	+2	&	[	&	29.5,	&	64.8\hspace{2pt}]	&	35.3	&	$\star$ \\
France	&	8	&	-1	&	[	&	50.8,	&	64.8\hspace{2pt}]	&	14.1	&	$\star\star\star$ \\
Luxembourg	&	9	&	-8	&	[	&	25.5,	&	58.9\hspace{2pt}]	&	33.4	&	$\star\star$ \\
Lithuania	&	10	&	+1	&	[	&	8.9,	&	57.1\hspace{2pt}]	&	48.2	&	$\star$ \\
Denmark	&	11	&	+2	&	[	&	16.7,	&	56.0\hspace{2pt}]	&	39.3	&	$\star$ \\
Sweden	&	12	&	+4	&	[	&	17.5,	&	55.3\hspace{2pt}]	&	37.8	&	$\star$ \\
Spain	&	13	&	-1	&	[	&	40.1,	&	54.8\hspace{2pt}]	&	14.7	&	$\star\star\star$ \\
Czechia	&	14	&	+1	&	[	&	14.7,	&	53.9\hspace{2pt}]	&	39.3	&	$\star$ \\
Poland	&	15	&	-5	&	[	&	23.3,	&	52.1\hspace{2pt}]	&	28.7	&	$\star\star$ \\
Hungary	&	16	&	+1	&	[	&	12.3,	&	49.6\hspace{2pt}]	&	37.3	&	$\star$ \\
Finland	&	17	&	-3	&	[	&	8.9,	&	47.2\hspace{2pt}]	&	38.3	&	$\star$ \\
Portugal	&	18	&	+3	&	[	&	8.5,	&	46.9\hspace{2pt}]	&	38.4	&	$\star$ \\
Croatia	&	19	&	-1	&	[	&	7.5,	&	45.6\hspace{2pt}]	&	38.1	&	$\star$ \\
Slovakia	&	20	&	+2	&	[	&	8.0,	&	45.6\hspace{2pt}]	&	37.6	&	$\star$ \\
Ireland	&	21	&	-2	&	[	&	1.7,	&	45.4\hspace{2pt}]	&	43.7	&	$\star$ \\
Malta	&	22	&	+3	&	[	&	12.0,	&	39.3\hspace{2pt}]	&	27.2	&	$\star\star$ \\
Latvia	&	23	&	-3	&	[	&	8.1,	&	38.2\hspace{2pt}]	&	30.2	&	$\star\star$ \\
Greece	&	24	&	+2	&	[	&	8.4,	&	36.8\hspace{2pt}]	&	28.4	&	$\star\star$ \\
Cyprus	&	25	&	-2	&	[	&	2.1,	&	36.6\hspace{2pt}]	&	34.5	&	$\star$ \\
Romania	&	26	&	-2	&	[	&	4.0,	&	36.6\hspace{2pt}]	&	32.6	&	$\star\star$ \\
Bulgaria	&	27	&	+1	&	[	&	4.1,	&	26.5\hspace{2pt}]	&	22.4	&	$\star\star$ \\
Estonia	&	28	&	-1	&	[	&	12.2,	&	17.3\hspace{2pt}]	&	5.1	&	$\star\star\star$ \\

\bottomrule

\end{tabular}

	}
	\label{tab:overall_but_ci}
\end{table*}

\section{Conclusion}

It is increasingly common to assess a multidimensional phenomenon through a composite indicator, and indeed the number of composite indicators in use internationally has been growing steadily. Surprisingly, there is no international agreement on their construction to date, and a different methodological approach accompanies each phenomenon. Appealing to simplicity in their construction,  a single number is expected to synthesise very complex phenomena without emphasising their limitations.

An important issue when constructing composite indicators comes from the choice of the aggregation methodology which involves the discussion about if compensability should or not be allowed. In this paper, we have proposed an alternative way to assess a multidimensional issue's performance by building a composite performance interval instead of getting only a real number. The MCPI allows for overcoming the criticism of composite indicators that they show a ``big-picture'' by providing a range of values. Making use of distance-based multicriteria techniques and different version of the Minkowski's $L_p$ metrics we have built two composite indicators which are  presented as a lower and upper bound of the MCPI. In this proposal, we have used the TOPSIS method primarily because of its ease of implementation in a wide variety of situations, with no restrictions on the number of alternatives or criteria. The lower bound corresponds to a non-compensability aggregation rule and provides the worst performance of the group of indicators or dimensions that has been aggregated. The upper bound is constructed allowing for full compensation using a linear rule based o weighted or additive aggregation. Finally, based on the balance property of the MCPI, we propose a $\beta$--Rating to identify those alternative in which there exist great MCPI span and, therefore, need to be analysed in greater depth to detect the dimension where action is needed with higher priority. 

As an example of application we compute the MCPI to assess the Circular Economy performance of European Member states by using the structure of indicators and dimensions provided by the European Commission Circular Economy monitoring framework. The overall Circular Economy MCPI has been calculated and also at the dimension level, thus providing a valuable tool to identify areas where countries need to concentrate their efforts to boost their circular economy performance. If we consider the four dimensions initially proposed in the circular economy theoretical framework, the ranking is headed by Luxembourg, although the dimensions are quite unbalanced. In the other hand, when the MCPI is constructed excluding the competitiveness and innovation dimension, the results seem more coherent, leading the ranking with the countries of Netherlands, Germany and Belgium with a better $\beta$--Rating.

Moving forwards, we see a promising line of future research in testing the applicability of the proposed MCPI by adopting other multicriteria techniques such as the reference point-based method or ELECTREE.  Besides, in the future, we plan to investigate if and how the MCPI could be applied to other domains in which composite indicators have been previously used, from sustainability to quality of life assessment, to cite but a few relevant applications.

\newpage


\clearpage\clearpage

\end{document}